# Relaxing Coherence Requirements on Laser Sources for Nanoscopy through Optical Fiber Technique

**KAIFA XIN[1,†], MENGDIE HOU[1,†], AO YANG[1,2,*], MÁRTON GELLÉRI[3], HYUN-KEUN LEE[1], ANTONIO VIRGILIO FAILLA[4], JIAFENG LU[5], XIANGLONG ZENG[6], MISCHA BONN[1], JESPER LÆGSGAARD[7] AND XIAOMIN LIU[1,*]**

*[1]Max Planck Institute for Polymer Research, Ackermannweg 10, 55128 Mainz, Germany*
*[2]School of Electrical and Information Engineering, Anhui University of Science and Technology, Huainan, 232001, China*
*[3]Institute of Molecular Biology (IMB), Ackermannweg 4, 55128 Mainz, Germany*
*[4]UKE Microscopy Imag-ing Facility (UMIF), University Medical Center Hamburg-Eppendorf (UKE), Martinistraße 52, 20246 Hamburg, Germany*
*[5]Institute of Fiber Optics, The Key Laboratory of Specialty Fiber Optics and Optical Access Network, Shanghai University, Shanghai 200444, China*
*[6]State Key Laboratory of Satellite Network, Shanghai Key Laboratory of Satellite Network, Shanghai Satellite Network Research Institute Co., Ltd, Shanghai 201210, China*
*[7]Department of Electrical and Photonics Engineering, Technical University of Denmark, ørsted Plads 343, 2800 Kgs. Lyngby, Denmark*
*[†]These authors contributed equally to this work.*
*[*]liuxiaomin@mpip-mainz.mpg.de*

**High numerical-aperture (NA) focusing of cylindrical vector beams (CVBs) typically requires costly, high-quality lasers to supply a stable vector pupil field. Here, we introduce fiber-conditioned vectorial nanofocusing, in which an optical-fiber-based mode-selective coupler projects a strongly distorted beam of a diode that costs two orders of magnitude less than the reference laser into radially or azimuthally polarized CVBs. A three-tolerance analysis clarifies its operating principle by establishing separate requirements for spatial state, temporal coherence and residual wavefront error. The resulting fields reproduce expected high-NA focal signatures and generate orientation-sensitive single-molecule excitation patterns consistent with reference-laser measurements, lowering the barrier to structured-light experiments.**

The ability to focus structured light (e.g., doughnut beams) through high-numerical-aperture (NA) objectives is a cornerstone of modern photonics. This capability enables applications from optical trapping[1–3] and far-field super-resolution microscopy[4–6] to near-field microscopy[7] and single-molecule dipole orientation measurements[8,9]. Meanwhile, it is also foundational for emerging quantum technologies, including quantum computing[10] and structured-light control of quantum materials[11]. The precise nanoscale structure of the focused electromagnetic field is critical in these regimes, as it governs light-matter interactions beyond the dipole approximation-a necessity now addressed by advanced ab initio theoretical frameworks[12].

The coherent benchmark for this process is the vectorial diffraction theory of Richards and Wolf[13,14], which maps a deterministic complex vector field at the objective pupil to the focal field by coherently superposing the angular-spectrum components. Conventionally, this pupil field is prepared using a premium laser, spatial filtering, and a precisely aligned free-space converter such as a q-plate, vortex retarder, or interferometric superposition of orthogonal modes[15], which must be centered on the beam axis. Once the pupil field is specified, however, the Richards-Wolf mapping is independent of how that field was prepared. The relevant design requirement therefore applies to the illumination system as a whole: it must deliver the intended vector state to the objective pupil. We therefore implement transverse-mode selection and vector-mode generation within the optical fiber-based delivery path, relaxing the requirement on the laser's transverse modal quality. This allows the source to be built around a semiconductor laser diode whose purchase price is two orders of magnitude below that of the commercial reference laser.

How closely such a system approaches the deterministic benchmark is governed by three distinct properties of the delivered field and focusing optics. Statistical fluctuations in amplitude, phase, and polarization modify the vector pupil state; finite source bandwidth can reduce the interference between pupil components associated with different optical paths; and residual wavefront error (WFE) introduces deterministic phase errors between them. Each mechanism perturbs the interference between pupil components from which the focus is formed. A common propagation description must therefore retain the mutual correlations of the vector field across the pupil at each optical frequency. These correlation are encoded in the pupil cross-spectral-density matrix (CSDM) $\mathbf{W}_p(\mathbf{s}_1, \mathbf{s}_2; \omega)$, which jointly describes spatial-coherence and polarization[16–18]. A deterministic system propagates the CSDM as $\mathbf{H}\mathbf{W}_p\mathbf{H}^\dagger$, and the detected intensity follows by spectrally integrating the trace of the focal CSDM. Under the narrowband and weakly dispersive conditions relevant here, the detected focal intensity for the CW fields takes the following compact pupil-pair form, which makes explicit the three factors governing focal-field formation:

$$I_{\rm det}(\mathbf{r}) \propto I_0 \iint_{\Omega_{\rm NA}^2} \sqrt{P_1 P_2}\ \mathrm{Tr}\left[\mathbf{G}_{0,1}\mathbf{M}_{p,12}\mathbf{G}_{0,2}^\dagger\right] \times \bar{\Gamma}_t\left(\Delta L_{12}^{(0)}\right)\exp\left[\mathrm{i}\frac{2\pi}{\lambda_0}\Delta W_{\rm ab,12}\right] d\Omega_1 d\Omega_2 \tag{1}$$

Here, $P_j$ is the normalized pupil intensity, $\mathbf{G}_{0,j}$ denotes the vectorial focusing kernel at $\lambda_0$, and $I_0$ collects the total pupil power and pupil-independent scalar factors. The normalized vector coherence matrix $\mathbf{M}_{p,12}$ describes the spatial-coherence and polarization correlations between pupil directions $\mathbf{s}_1$ and $\mathbf{s}_2$ . The temporal-coherence factor $\bar{\Gamma}_t\left(\Delta L_{12}^{(0)}\right)$ accounts for spectral averaging of the interference between two pupil directions with optical-path difference $\Delta L_{12}^{(0)}$, while $\Delta W_{\rm ab,12}$ is their differential residual WFE. The full derivation and reduction conditions are provided in Supplementary Section S2.

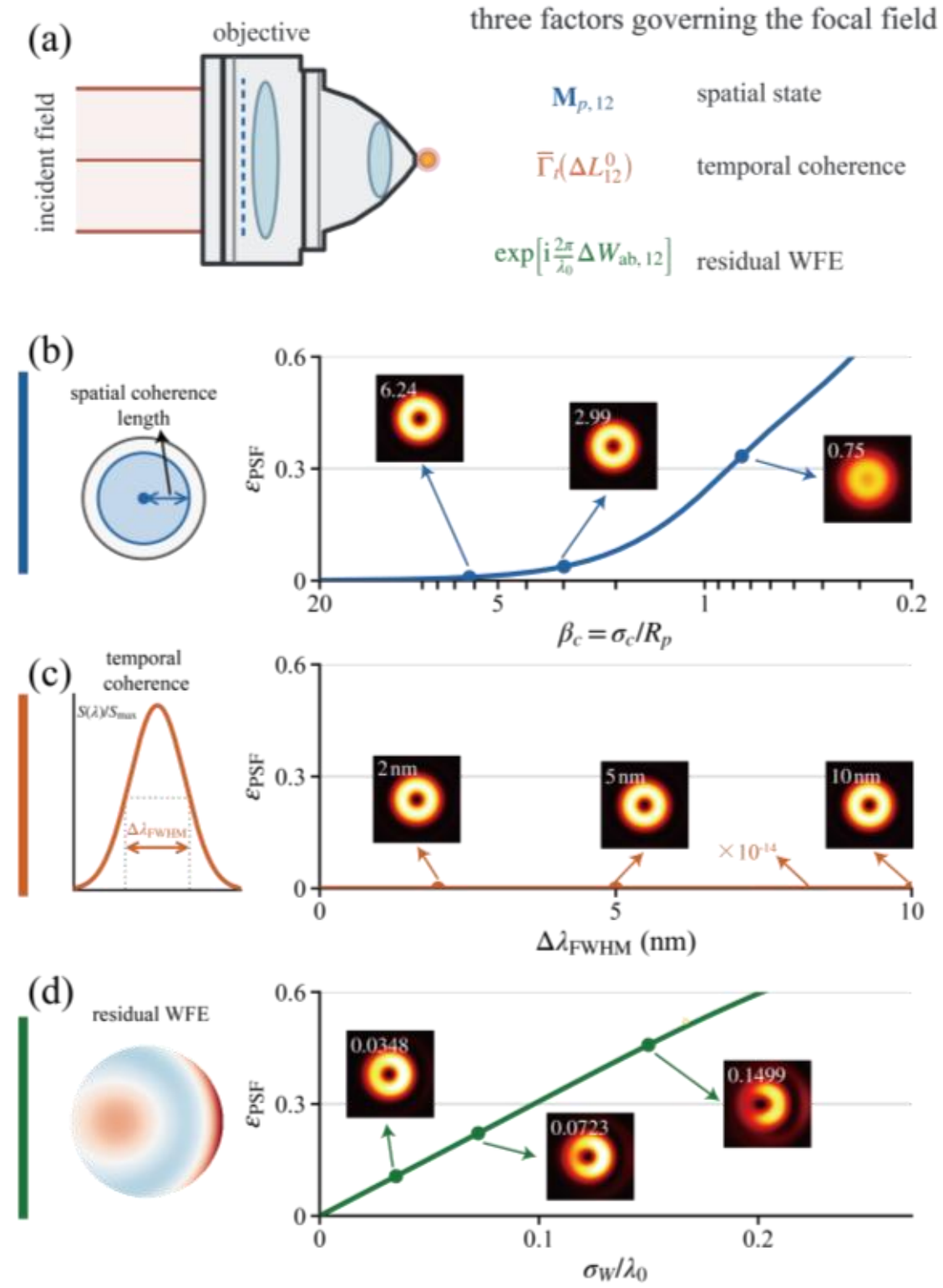

Fig. 1. Three constraints governing high-NA vectorial focusing. (a) An aplanatic objective maps the pupil field onto the focus. Equation (1) shows the three factors degrade the ideal focus: the normalized vector coherence matrix $M_{p,12}$ (spatial state), the temporal-coherence factor $\bar{\Gamma}t\left(\Delta L_{12}^{(0)}\right)$ (temporal coherence), and the residual wavefront phase $\exp[\mathrm{i}2\pi\Delta W_{\rm ab,12}/\lambda_0]$ (WFE). (b-d) Each factor is degraded alone, with the other two kept at their ideal values, for an azimuthally polarized CVB ($TE_{01}$) at $\lambda_0$ = 647nm and NA = 1.40. In each row, the left panel illustrates the pupil quantity that is varied, and the right panel shows the resulting relative PSF deviation $\epsilon_{\rm PSF}$ from the ideal focus; The varied quantity is the Gaussian–Schell coherence width $\beta_c = \sigma_c / R_{\rm p}$ in (b), the source bandwidth $\Delta\lambda_{FWHM}$ in (c), and the RMS wavefront error $\sigma_W/\lambda_0$ in (d). The three vertical scales are identical, and only (c) remains flat. The reference in each case is the fully coherent, monochromatic, and aberration-free focus. Definitions and settings are given in Supplement Section S3.

Figure 1 distinguishes the practical roles of the three factors in Eq. (1). Under the adopted aplanatic, weakly dispersive model, increasing the Gaussian spectral bandwidth to ~10 nm produces a negligible change in the focal intensity, as shown in Fig. 1(c). Measurements before and after MSC conversion establish coherence-length lower bounds of 1 m for both sources. These values exceed the maximum pupil-pair differential OPD of the focusing system by at least five orders of magnitude, confirming that temporal coherence is not limiting in the present narrowband CW implementation. Residual WFE can appreciably distort the focus, as shown in Fig. 1(d), but it is a deterministic system error that can be characterized and reduced through wavefront correction. Figure 1(b) shows that degradation of the pupil state fills the central null and redistributes the annular focal intensity. Preserving the target focal structure therefore requires the delivered pupil field to be dominated by a single spatially coherent mode that reproduces the desired complex vector distribution. In the normalized-CSDM description, the dominant coherent mode $\psi_0$ must carry a pupil-power fraction ($f_0 \geq 1 - \varepsilon_{\rm rank}$) and have a target-mode fidelity ($F_{\rm tar} \geq 1 - \varepsilon_{\rm mode}$).

Both requirements on spatial state are addressed by the practical fiber-based solution shown in Fig. 2, in which pupil-state preparation is transferred from the laser source to the fiber delivery path.

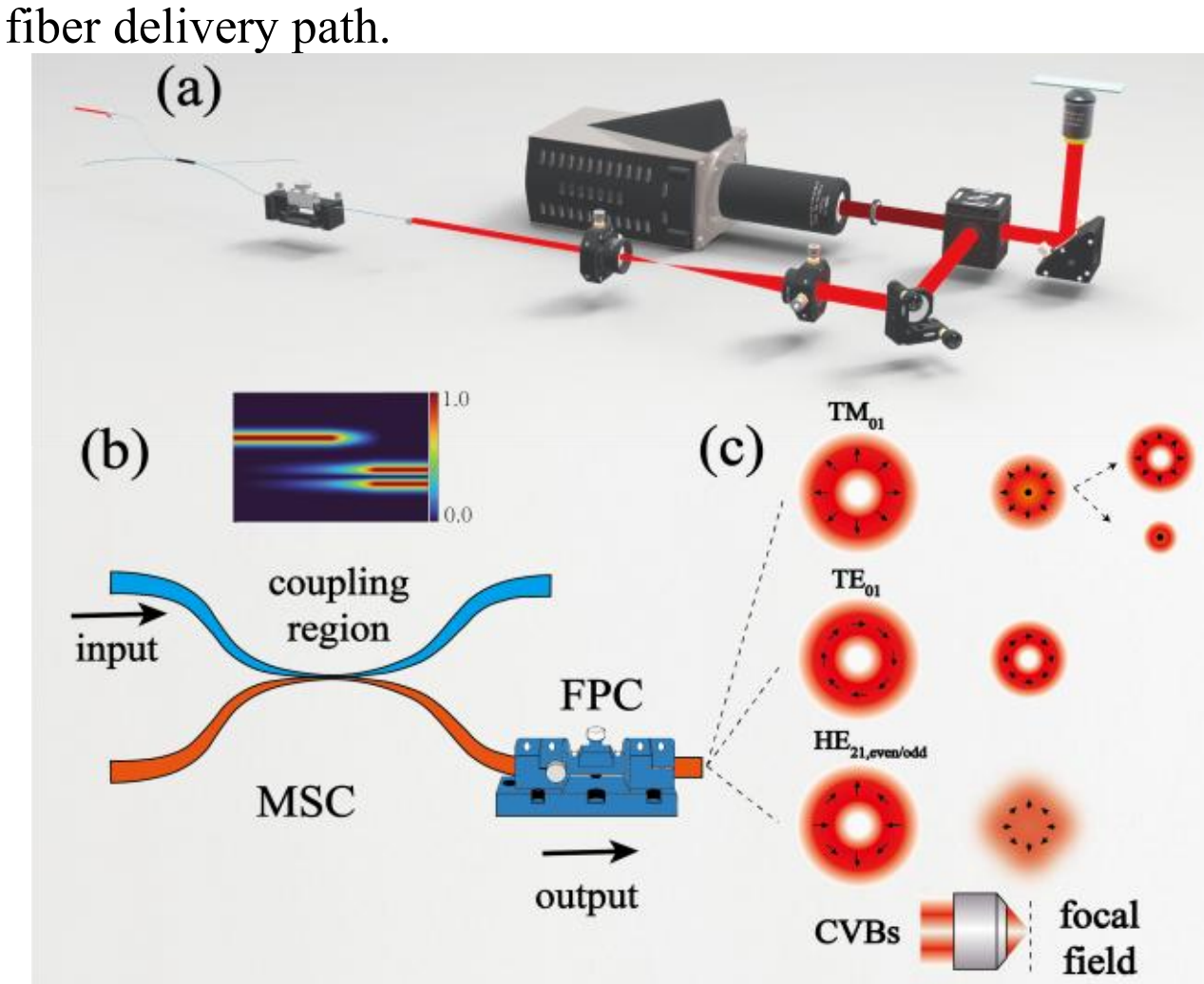

Fig. 2. Single-molecule orientation microscope using MSC-generated CVBs. (a) Experimental setup. Continuous-wave light is projected onto a single guided mode, converted by the mode-selective coupler (MSC), collimated, expanded, and focused by a high-NA objective onto TDI molecules immobilized on a coverslip; a three-axis

piezoelectric stage raster-scans the sample through the focus. (b) MSC principle: the coupler is phase-matched between $LP_{01}$ mode in the input fiber and the $LP_{11}$ family in the output fiber, and the fiber polarization controller (FPC) selects the vector eigenmode within that family. Inset, simulated power evolution in the coupling region. (c) Calculated fiber-output and focal fields of the $TM_{01}$, $TE_{01}$ and even/odd $HE_{21}$ modes. carries a strong on-axis longitudinal component $E_z$ at the focal field.

The architecture satisfies the two pupil-state requirements sequentially. The input fiber and mode stripper project the usable source power onto the guided $LP_{01}$ channel, establishing a single spatially coherent mode. Phase-matched coupling then transfers the accepted field into the $LP_{11}$ family, from which the FPC selects the target $TE_{01}$ or $TM_{01}$ vector mode. As deterministic transformations, the latter operations preserve the fixed-frequency CSDM rank while setting the required vector distribution. The source transverse structure therefore primarily determines the coupling efficiency, whereas the fiber system determines the spatial state delivered to the high-NA microscope.

With the operating principle established, the performance of the fiber-conditioned CVBs is evaluated across the complete experimental chain. The OBIS reference laser and the strongly distorted home-assembled (HA) diode source are compared first through their polarization-resolved output fields and bead-mapped high-NA focal responses in Fig. 3, and then through orientation-sensitive single-molecule excitation in Fig. 4. Together, these measurements test whether the spatial state prepared by the fiber system, rather than the native transverse profile of the source, governs the performance relevant to vectorial nanofocusing.

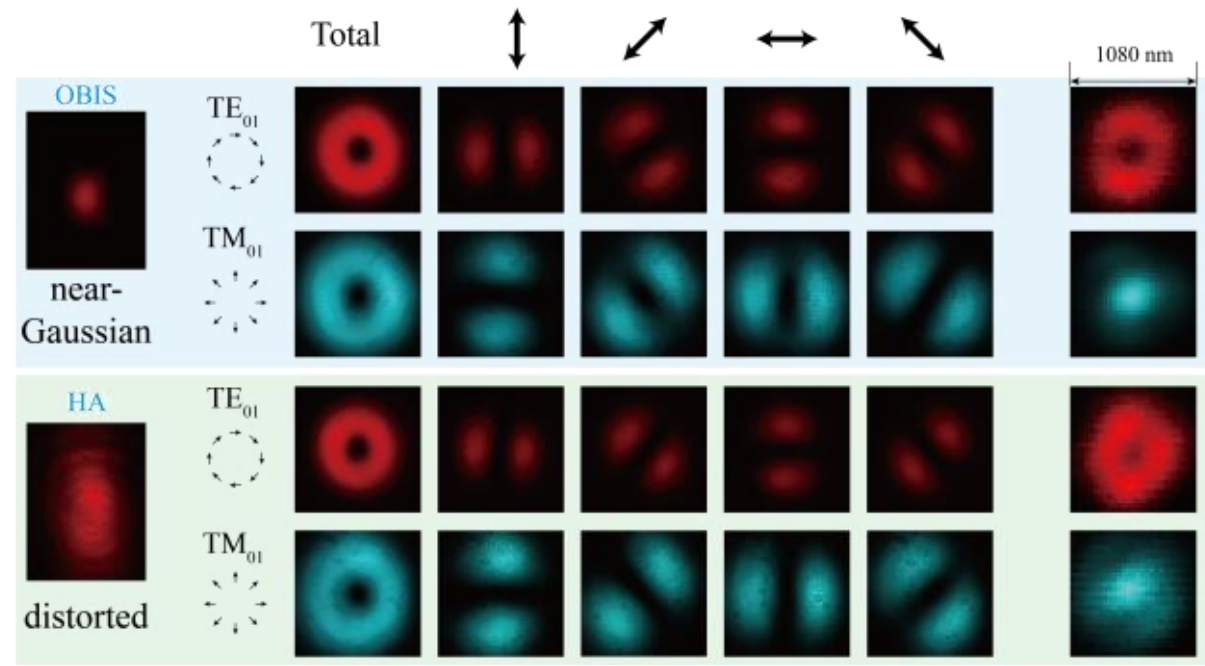

Fig. 3. CVBs generated by the MSC from a reference and a low-cost source. (a) Collimated input intensity profiles of the OBIS reference laser, which is near-Gaussian, and of the HA diode laser, which is strongly distorted. (b) Polarization-resolved output for each source and target mode. The first column is the total intensity, and the remaining columns correspond to the analyzer transmission axes indicated above. The lobe pattern rotates with the analyzer axis, perpendicular to it for $TE_{01}$ and parallel to it for $TM_{01}$, as expected for azimuthal and radial polarization. (c) Focal responses obtained by scanning 40 nm fluorescent beads through the focus of an NA = 1.40 objective; the indicated width is 1080 nm.

Despite the markedly different input profiles in Fig. 3(a), both sources produce the characteristic polarization-resolved signatures of the $TE_{01}$ and $TM_{01}$ modes after MSC conversion, as shown in Fig. 3(b). Modal decomposition gives estimated target-mode weights above 91% for all four source-mode combinations. The bead scans in Fig. 3(c) further recover the expected annular $TE_{01}$ and centrally peaked $TM_{01}$ focal responses for both sources, establishing the expected focal topology after high-NA focusing. Going beyond these intensity-based measurements, an immobilized molecule with a well-defined linear transition dipole can serve as a local, orientation-sensitive probe of the vectorial focal field. Within the electric-dipole approximation, its excitation probability is described by $p = |\vec{\mu} \cdot \vec{e}|^2$, where $\vec{\mu}$ is the molecular transition dipole moment and $\vec{e}$ is the local electric field. We implement this test using individual TDI molecules raster-scanned through the focused $TE_{01}$ and $TM_{01}$ fields generated from the OBIS and HA sources. Because $\vec{\mu}$ remains fixed during each scan, the resulting molecular scan maps the coupling between the vectorial focal field and the absorption transition dipole of an individual molecule.

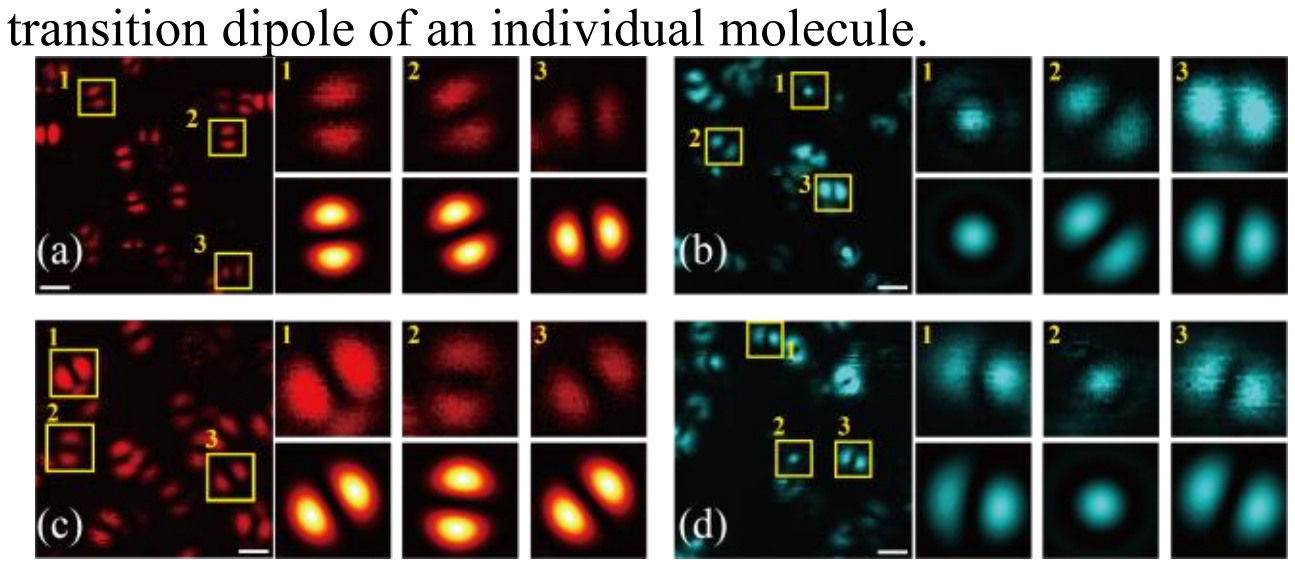

Fig. 4. Orientation-sensitive excitation patterns of individual TDI molecules obtained with MSC-generated CVBs. (a, b) Fluorescence raster-scan images acquired using the reference OBIS laser after conversion into the $TE_{01}$ (a) and $TM_{01}$ (b) modes. (c, d) Corresponding images acquired using the HA diode laser after conversion into the $TE_{01}$ (c) and $TM_{01}$ (d) modes. In each panel, the overview image shows isolated immobilized TDI molecules, with three representative molecules marked by numbered yellow boxes. The upper row of adjacent images shows enlarged experimental patterns from the corresponding boxes, whereas the lower row shows representative patterns calculated using vectorial diffraction theory and electric-dipole coupling. Both sources produce the expected families of orientation-dependent responses, including directional two-lobed patterns under $TE_{01}$ excitation and centrally weighted or directional patterns under $TM_{01}$ excitation. White scale bar, 1 µm

The overview scans in Fig. 4 contain isolated molecules exhibiting a variety of orientation-dependent excitation patterns. Under $TE_{01}$ excitation, the selected molecules show directional two-lobed patterns with different lobe axes, as expected from coupling to the transverse azimuthal electric field. Under $TM_{01}$ excitation, both centrally weighted and directional two-lobed responses are observed, consistent with the contributions of the longitudinal and radial electric-field components to molecules with different dipole orientations. Similar families of excitation patterns are observed in the independent OBIS and HA data sets and are consistent with previous single-molecule studies using tightly focused radial and azimuthal polarizations. The calculated excitation patterns shown below the measurements were generated for selected molecular dipole orientations using the modeled vectorial focal fields and the coupling relation defined above. They illustrate the characteristic morphologies expected under $TE_{01}$ and $TM_{01}$ excitation and are qualitatively consistent with the measured patterns. Together with the high target-mode weights and the bead-mapped focal responses in Fig. 3, these results show that the low-cost diode source supports orientation-sensitive single-molecule excitation comparable to that obtained with the single-transverse-mode reference laser.

Taken together, these results support a specific conclusion: the vector field required for high-NA vectorial focusing need not originate from the laser, because the governing condition applies to the field delivered at the objective pupil and can be established by an engineered delivery system. The delivery architecture can therefore relax the spatial-state requirement placed on the source. More broadly, the three-tolerance framework serves not only as an analytical model but also as an instrument-design procedure. Because the three tolerances arise from physically distinct mechanisms, they should be evaluated and verified independently rather than collapsed into a single source specification. For each experiment, their allowable ranges must be determined from the target pupil field, system OPDs, residual WFEs, and chosen performance metric. This separate evaluation reveals both the performance-limiting elements and the parts of the system in which lower-cost components can be used without compromising the required performance.

The most immediate extension is full fiber integration. In the present system, the only remaining free-space step is launching the source into the input arm. Connecting a pigtailed diode to this arm through a fiber connector would remove this alignment and leave no adjustable free-space element before CVB generation. Together with automated modal control, this configuration would reduce source exchange to reconnecting a fiber and enable comparison experiments. An input fiber that is intrinsically single-mode at the operating wavelength would further make the initial spatial-state preparation deterministic, rather than relying on loss-based rejection of unwanted modes through mode stripper.

Pulsed operation is a more demanding but plausible extension. The broad MSC conversion bandwidth[19] provides a route for extending the present high-NA architecture to pulsed operation, building on previous MSC-based generation of pulsed CVBs and femtosecond vortex beams[20,21]. This extension would require sufficiently uniform conversion efficiency, target-mode composition, and spectral phase across the pulse spectrum. More broadly, fiber-conditioned vector fields could be adapted to MINFLUX nanoscopy[5,6], optical trapping and manipulation[1,2], and near-field excitation[7].

**Funding.** This project has received funding from the European Union's Horizon Europe Research and Innovation Programme under the Marie Sklodowska-Curie Grant Agreement No. 101072409(HOMTech). This work in Mainz, Germany was partially financially supported by the Max Planck Society and the Deutsche Forschungsgemeinschaft (DFG, German Research Foundation) in the framework of the collaborative research center "Defects and Defect Engineering in Soft Matter" (SFB1552) under Project No. 465145163. M. Hou acknowledges support from the Chinese Scholarship Council (CSC).

**Acknowledgment.** We thank Qiqi Yang for preparing the TDI samples, Xingfu Zhu for assistance with the preparation of the fluorescent bead samples. We are particularly grateful to Marc Jan Zadel and Florian Gericke for their invaluable technical support throughout this work, including discussions on experimental hardware design, assistance with constructing the experimental setup, and the design and fabrication of custom components. We also thank Prof. Christoph Cremer for valuable discussions.

**Author's contributions.** Kaifa Xin and Mengdie Hou contributed equally to this work and jointly developed the theoretical framework. Kaifa Xin coordinated the project, designed and performed the experiments, analyzed the experimental data, prepared the figures, and wrote the original draft. Mengdie Hou performed the numerical simulations, contributed to data analysis, and participated in manuscript preparation. Ao Yang and Hyun-Keun Lee contributed to the early development of the project and its experimental methodology. Márton Gelléri and Antonio Virgilio contributed to scientific discussions throughout the experimental work. Jiafeng Lu and Xianglong Zeng contributed to the design and fabrication of the mode-selective coupler and to the interpretation of its performance. Jesper Lægsgaard contributed to the fiber-mode analysis and theoretical interpretation. Mischa Bonn and Xiaomin Liu supervised the project. Xiaomin Liu conceived and directed the study. All authors discussed the results, reviewed the manuscript, and approved its final version.

**Disclosures.** X. Kai, M. Hou, J. Lu, X. Zeng, M. Bonn and X. Liu are listed as inventors of one patent application related to the work presented in this manuscript. All other authors have nothing to disclose.

**Data availability.** Data underlying the results presented in this paper are not publicly available at this time but may be obtained from the authors upon reasonable request.

## References

1. M. Padgett and R. Bowman, Nat. Photonics **5**, 343 (2011).
2. A. Kritzinger, A. Forbes, and P. B. C. Forbes, Sci. Rep. **12**, 17690 (2022).
3. K. Dholakia and T. Čižmár, Nat. Photonics **5**, 335 (2011).
4. S. W. Hell, Science **316**, 1153 (2007).
5. F. Balzarotti, Y. Eilers, K. C. Gwosch, A. H. Gynnå, V. Westphal, F. D. Stefani, J. Elf, and S. W. Hell, Science **355**, 606 (2017).
6. K. C. Gwosch, J. K. Pape, F. Balzarotti, P. Hoess, J. Ellenberg, J. Ries, and S. W. Hell, Nat. Methods **17**, 217 (2020).
7. J. Zeng, F. Huang, C. Guclu, M. Veysi, M. Albooyeh, H. K. Wickramasinghe, and F. Capolino, ACS Photonics **5**, 390 (2018).
8. H. Ishitobi, I. Nakamura, N. Hayazawa, Z. Sekkat, and S. Kawata, J. Phys. Chem. B **114**, 2565 (2010).
9. L. Novotny, M. R. Beversluis, K. S. Youngworth, and T. G. Brown, Phys. Rev. Lett. **86**, 5251 (2001).
10. S. J. Evered, D. Bluvstein, M. Kalinowski, S. Ebadi, T. Manovitz, H. Zhou, S. H. Li, A. A. Geim, T. T. Wang, N. Maskara, H. Levine, G. Semeghini, M. Greiner, V. Vuletić, and M. D. Lukin, Nature **622**, 268 (2023).
11. D. Session, M. Jalali Mehrabad, N. Paithankar, T. Grass, C. J. Eckhardt, B. Cao, D. Gustavo Suárez Forero, K. Li, M. S. Alam, K. Watanabe, T. Taniguchi, G. S. Solomon, N. Schine, J. Sau, R. Sordan, and M. Hafezi, Nat. Photonics **19**, 156 (2025).
12. F. P. Bonafé, E. I. Albar, S. T. Ohlmann, V. P. Kosheleva, C. M. Bustamante, F. Troisi, A. Rubio, and H. Appel, Phys. Rev. B **111**, 085114 (2025).
13. E. Wolf, (n.d.).
14. B. Richards and E. Wolf, Proc. R. Soc. Lond. Ser. Math. Phys. Sci. **253**, 358 (1959).
15. Q. Zhan, Adv. Opt. Photonics **1**, 1 (2009).
16. E. Wolf, Phys. Lett. A **312**, 263 (2003).
17. J. Tervo and J. Turunen, Opt. Commun. **209**, 7 (2002).
18. J. Tervo, T. Setälä, and A. T. Friberg, JOSA A **21**, 2205 (2004).
19. H. Yao, F. Shi, Z. Wu, X. Xu, T. Wang, X. Liu, P. Xi, F. Pang, and X. Zeng, Nanophotonics **9**, 973 (2020).
20. T. Wang, F. Wang, F. Shi, F. Pang, S. Huang, T. Wang, and X. Zeng, J. Light. Technol. **35**, 2161 (2017).
21. H. Wan, J. Wang, Z. Zhang, Y. Cai, B. Sun, and L. Zhang, Opt. Express **25**, 11444 (2017).